\documentclass[fleqn,10pt]{olplainarticle}
\usepackage{setspace}

\title{Anomalous DNA hybridisation kinetics on gold nanorods revealed via a dual single-molecule imaging and optoplasmonic sensing platform}

\author[1,2,+,*]{Narima Eerqing}
\author[1,2,+]{Hsin-Yu Wu}
\author[1,2]{Sivaraman Subramanian}
\author[1,2]{Serge Vincent}
\author[1,2]{Frank Vollmer}
\affil[1]{Living Systems Institute, University of Exeter, Stocker Road, Exeter EX4 4QD, United Kingdom}
\affil[2]{Department of Physics and Astronomy, University of Exeter, Stocker Road, Exeter EX4 4QL, United Kingdom}
\affil[+]{$These$ $authors$ $have$ $contributed$ $equally$ $to$ $this$ $work$ $and$ $share$ $first$ $authorship$}
\affil[*]{$Email: ne276@exeter.ac.uk$  /  $Tel:0044$ $07840355637$}


\begin{abstract}
Observing the hybridisation kinetics of DNA probes immobilised on plasmonic nanoparticles is key in plamon-enhanced fluorescence detection of weak emitting species, and refractive index based single-molecule detection on optoplasmonic sensors. The role of the local field in providing plasmonic signal enhancements for single-molecule detection has been studied in great detail. Nevertheless, few studies have compared the experimental results in both techniques for single-molecule studies. Here we developed the first optical setup that integrates optoplasmonic and DNA-PAINT based detection of oligonucleotides to compare these sub-platforms and provide complementary insights into single-molecule processes. We record the fluorescence and optoplasmonic sensor signals for individual, transient hybridisation events. The hybridisation events are observed in the same sample cell and over a prolonged time (i.e. towards high binding site occupancies). A decrease in the association rate over the measurement duration is reported. Our dual optoplasmonic sensing and imaging platform offers insight into the observed phenomenon, revealing that irreversible hybridisation events accumulate over detected step signals in optoplasmonic sensing. Our results point to novel physicochemical mechanisms that result in the stabilisation of DNA hybridisation on optically-excited plasmonic nanoparticles.
\end{abstract}
\begin{document}

\flushbottom
\maketitle 
\thispagestyle{empty}

\section*{Introduction}

The interaction between DNA strands is key to many fundamental processes in the cell. The hybridisation between DNA oligonucleotides is essential for our most sensitive methods of DNA detection, including state-of-the-art single-molecule     techniques\cite{baaske2014single,eerqing2021comparing,zhang2018predicting}. Single-molecule techniques have enriched biomolecular studies by providing details about the kinetics of biological reactions and physiological processes that is not apparent in their corresponding bulk measurements. Powerful new approaches to single-molecule sensing and imaging have emerged in the last few decades. One example is fluorescence based single-molecule imaging, which overcomes the diffraction limit by reconstructing images from high-precision temporal modulation and the accumulation of single-molecule detection events\cite{li2018switchable,khater2020review,jradi2019chemistry,diekmann2020optimizing}. Among these, photo-activated localisation microscopy (PALM) \cite{owen2010palm,bouzin2019photo}, stochastic optical reconstruction microscopy (STORM) \cite{rust2006sub,xu2017stochastic}, and DNA based point accumulation for imaging in nanoscale topography (DNA-PAINT) \cite{schnitzbauer2017super,strauss2020up,filius2020high,lin20203d} have robustly demonstrated single-molecule localisation microscopy at the nanoscale. On the other hand, noble metal nanoparticles of various morphologies have drawn attention for their use in single-molecule sensing due to their extraordinary optical properties derived from localised surface plasmon resonances (LSPRs). In a large range of applications, plasmonic nanoparticles have been employed to amplify single-molecule detection signals. Examples for this are the plasmonic enhancement of fluorescence signals adjacent to nanoparticles, \cite{bardhan2009fluorescence,paulo2018enhanced,yuan2013thousand} and the enhancement of the label-free signals from whispering-gallery mode (WGM) sensors \cite{subramanian2022optoplasmonic,baaske2014single,subramanian2021sensing,eerqing2021comparing,vollmer2020optical,bowen2022single,yu2022single}.

Along with the development of various single-molecule techniques, it is becoming increasingly important to compare and cross-validate their results \cite{bowen2022single}. Detecting a single-molecule process on two different optical instruments enables one to gain a deeper understanding of the biomolecular system under investigation. Recently, we reported a study in which we compare DNA hybridisation events observed on plasmonic nanorods using an optoplasmonic sensor, with the results obtained on a single-molecule imaging technique based on DNA-PAINT \cite{eerqing2021comparing}. The optoplasmonic sensor measures single-molecule events within the enhanced near field of plasmonic gold nanorods (GNRs) that are attached to an optical WGM surface. The signals are obtained indirectly via the shift in the resonance of the WGM. On the other hand, DNA-PAINT provides signals via fluorescence localization microscopy. Although DNA-PAINT does not require plasmonic enhancement, we performed all DNA-PAINT experiments with molecular interactions on the surface of GNRs to replicate the conditions of the optoplasmonic sensor system. We found that both techniques deliver comparable results. Specifically, we found DNA dissociation kinetics (i.e. off-rates) for both schemes lay within experimental error.

In this article, we demonstrate the first use of a total internal reflection fluorescence (TIRF) objective to perform label-free optoplasmonic sensing and fluorescence imaging of single molecules in one optical platform. The TIRF objective is employed as an evanescent coupler similar to common coupling methods such as prism, grating, end-fibre and wave guide couplers to evanescently excite the WGMs on glass microspheres while enabling single-molecule imaging capability. We use this platform to study the hybridisation kinetics of DNA oligomers attached to gold nanorods (GNRs). We chose to study the interaction between DNA oligonucleotides because recent reports show that DNA hybridisation kinetics on GNRs are seemingly affected by the experimental single-molecule technique.  In 2018, Weichun \textit{et al.}\cite{zhang2018single} studied the single-molecule fluorescence enhancement from GNRs, reporting a disappearance of DNA hybridisation events over time. This phenomenon was, however, not observed for 'docking' strands (immobilized single stranded DNA) bound to glass. Previously, Taylor \textit{et al.}\cite{taylor2018all} conducted DNA-PAINT experiments on gold nanorods to reconstruct single GNR geometry. They also reported a similar reduction and disappearance of DNA hybridisation events for 'docking' strands attached to GNRs.  Weichun \textit{et al.}\cite{zhang2018single} attributed this effect to the cleaving of $Au-S$ bonds and thus removal of the docking DNA strands bound to the GNRs by hot electrons generated in the GNRs. Studying the decrease in DNA hybridisation event frequency on our dual single-molecule fluorescence imaging and optoplasmonic sensing platform would provide more detailed insight into the mechanisms behind the reported phenomenon. By probing single-molecule kinetics with fluorescence and  optoplasmonic refractive index based methods in parallel, we obtain results which show that the anomalous disappearance of DNA hybridisation events over time on GNR arises from the probabilistic permanent binding of complementary strands and that this is not a result of hot electron cleaving of $Au-S$ bonds.

\section*{Dual sensing and imaging setup}

\begin{figure}[t!]
\centering
\includegraphics[width=0.98\linewidth]{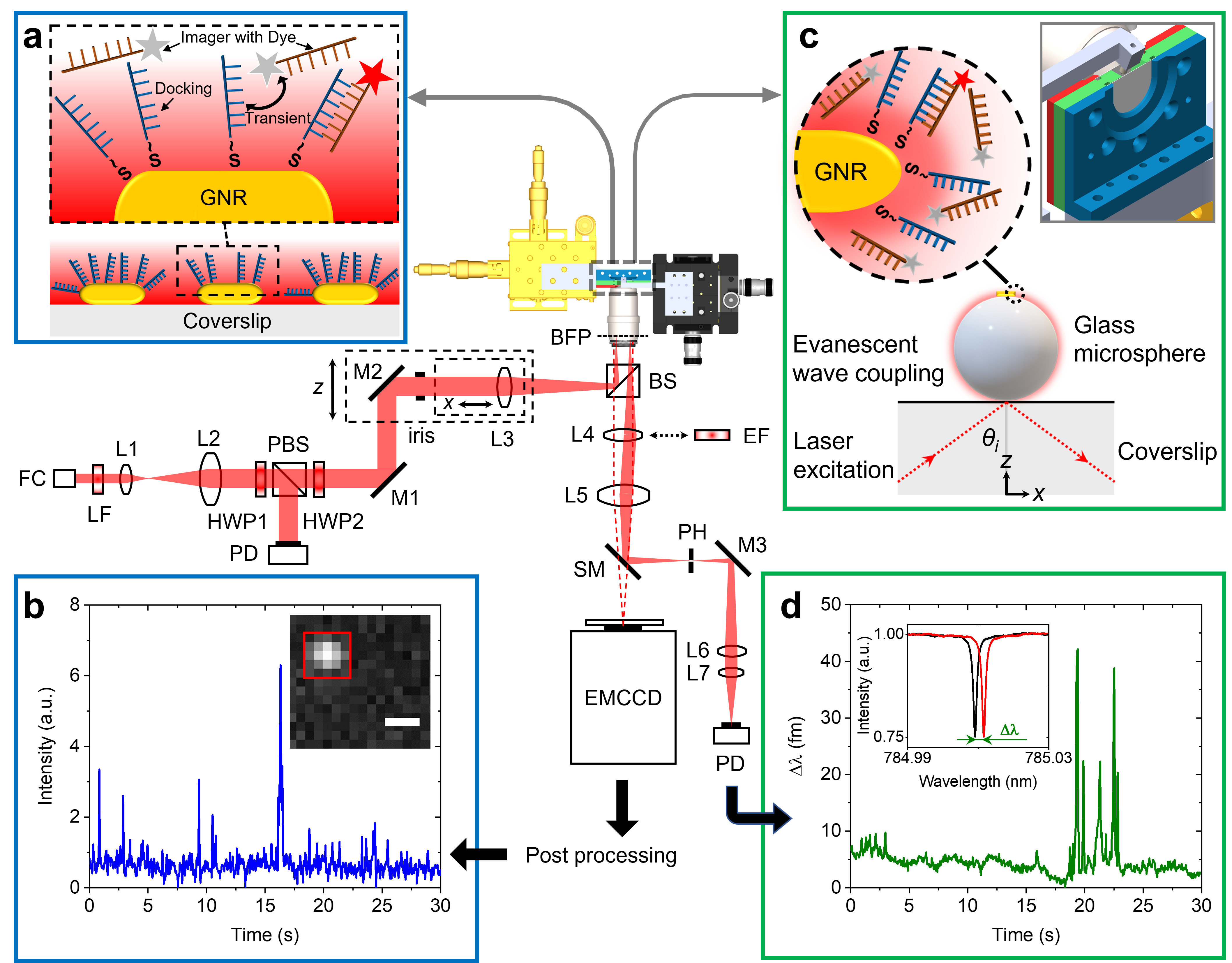}
\caption{Design of the dual optoplasmonic sensing and imaging platform. Incident laser beam is focused onto the back focal plane of a TIRF objective, establishing total internal reflection on the glass coverslip surface. Generated evanescent wave is then coupled into the WGM. (a) Schematic of DNA-PAINT imaging. The zoomed-in view shows the transient interaction between freely defusing imager DNA strands and docking DNA strands immobilised on the GNRs. The imager denoted by a red star is the one that contributes fluorescence signals. (b) Typical intensity time trace extracted from colocalised events in the 5$\times$ 5 pixels (with red ROI box in the inset) around the GNR. The scale bar = 500 nm. (c) Schematic describing the optoplasmonic sensing principle. The zoomed-in view shows imager strands interact with docking strands, wherein the imager denoted by a red star is the molecule that contributes to the sensing signal. The inset depicts the chamber and microsphere positioning. (d) Typical WGM resonance wavelength trace obtained by tracking the resonance peak position (with Lorentzian lineshape shown in the inset). FC: fiber-coupled collimator; LF: laser-line filter; L: lens; HWP: half-wave plate; PBS: polarizing beam splitter; M: Mirror; BS: non-polarizing beam splitter; BFP: back focal plane; EF: emission filter; SM: switchable mirror; PD: photodetector.}

\label{fig:1}
\end{figure}

Figure\,\ref{fig:1} shows the schematic of the experimental setup that we have developed for dual single-molecule imaging microscopy and optoplasmonic sensing (SIMOPS). Light from a tunable external cavity diode laser (Toptica DL pro 780) is collimated, expanded, and focused onto the back focal plane of a TIRF objective (CFI Apo TIRF 100X, 1.49 NA, Nikon). The total internal reflection (TIR) off a glass coverslip is used to evanescently couple to optical WGMs of a fused silica microsphere similar to evanescent coupling via prisms\cite{baaske2014single}. The reflected light is then collimated and collected by a lens L4, and a switchable mirror (SM) onto a photodetector (PD). Alternatively, the evanescent field originating from TIR off the glass coverslip is used to excite plasmonic GNRs and dye molecules directly near the surface of the glass coverslip. (Please refer to Supplementary Figure S1 for details of field enhancements.) In this case, a fluorescence image is  obtained by replacing the lens L4 with a switchable emission filter and collecting the emitted photons on an EMCCD camera (Andor iXon 888). 

Figure\,\ref{fig:1}a shows a schematic of the DNA-PAINT measurements. Firstly, plasmonic GNRs are immobilised on the surface of a glass coverslip that is placed in the sample chamber. The GNR attachment can be monitored in real-time via the photoluminescence of the GNRs as shown in Figure\,\ref{fig:1}b inset (highlighted by the red box). Secondly, DNA docking strands are then anchored onto the GNRs via mercaptohexyl linkers. The sample chamber is then rinsed to remove excess, unbound DNA strands. Finally, complementary DNA strands with a fluorescent label (DY782) are added to the sample chamber (See Materials and Methods for more details). The transient hybridisation of freely-diffusing imager strands and fixed docking strands then produces an increased intensity at the GNR location. The intensity integrated within a region-of-interest (ROI) of 5 $\times$ 5 pixels around the GNR position then provides the intensity time traces as shown in Figure\,\ref{fig:1}b. The captured ROIs are stacked and processed in Fiji (ImageJ) via a single-molecule localisation microscopy package named ThunderSTORM\cite{ovesny2014thunderstorm}. 

Figure\,\ref{fig:1}c shows the schematic of the experiments performed using the optoplasmonic technique. In this case, light is coupled to WGMs in an \SI{80}{\micro m} glass microsphere\cite{subramanian2022optoplasmonic} placed near the coverslip surface. GNRs are then attached to the microsphere surface. Subsequently, docking DNA strands are immobilised on the GNRs with a protocol similar to that for the DNA-PAINT. Once the fluorophore-labeled imager strands are added to the sample chamber (Detailed protocol can be found in the Materials and Methods section), hybridisation of the imager strands with the docked DNA strands is observed in the form of a shift in WGM resonance frequency (See inset of Figure\,\ref{fig:1}d ). These frequency shifts are recorded over time to obtain the single-molecule time traces as shown in Figure\,\ref{fig:1}d (See Materials and Methods for further context).

\section*{Materials and methods}
\subsection*{Materials}

GNRs with an average diameter of 10 nm and length of 35 nm (i.e. longitudinal plasmon resonance at $\lambda \approx$ 750 nm) were purchased from Nanopartz Inc.(A12-10-CTAB-750).  All DNA oligos were purchased from Eurofins Genomics and their sequences are listed in Table 1. Glass coverslips with a refractive index of n $\approx$ 1.52, with an aspect ratio of 22 mm $\times$ 22 mm and a thickness of 170 $\pm$ \SI{5}{\micro m}, were purchased from Thorlabs (Precision Glass Cover Slips, CG15CH). High-$Q$ glass microspheres (n$\approx$ 1.45) were fabricated with a 30 W CO$_{2}$ continuous wave laser ($\lambda$ $\approx$ \SI{10.6}{\micro m}) purchased from Synrad 48-2, Novanta Inc., WA, USA.  PLL-g-PEG was purchased from SuSOS Surface Technology, Switzerland.  Tris(carboxyethyl)phosphine hydrochloride (TCEP)  was purchased from Sigma-Aldrich (Catalog Number 646547). 4-(2-hydroxyethyl)-1-piperazineethanesulfonic acid (HEPES) was purchased from Sigma Aldrich, and a 20 mM HEPES buffer with pH$\approx$7 was prepared for use as the interaction buffer. All chemicals for the buffers were purchased from Sigma Aldrich.

\begin{table}[hb]
\renewcommand{\arraystretch}{1.5}
\centering
 \caption{Sequences of ssDNA used for the experiments.}
\label{tab:T1}
\begin{tabular}{lll}
\hline
& \textbf{ssDNA}   & \textbf{Sequence (5'-3') }\\
\hline
\multirow{3}{4em}{Set I} & P1               & [ThiolC6] TTT TAT ACA TCT A          \\ 
                         & ImP1*D           & [DY782] CTA GAT GTA T                 \\
    
\hline
\multirow{3}{4em}{Set II}& T22              & [ThiolC6] TTT TGA GAT AAA CGA GAA GGA TTG AT \\
                 & ImT22*D           & [DY782] ATC AGT CCT TTT CCT TTA TCT C  (3 mismatched)  \\
\hline
\end{tabular}
\end{table}

\subsection*{SIMOPS Setup Optics}

Figure 1\,\ provides the schematic of the SIMOPS setup. In order to couple light into WGMs with an objective lens, a fiber-coupled tunable laser diode with nominal wavelength of 780 nm (DL pro, Toptica) was collimated with a fiber collimator and then expanded by a beam expander (L1 and L2), resulting in a collimated beam diameter of 17.3 mm. A half-wave plate (HWP1) and a polarizing beam splitter (PBS; Tp:Ts $>$ 3000:1) were used to control the intensity of the p-polarised transmitted beam. An additional half-wave plate (HWP2) was placed behind the PBS to alter the polarisation state of the transmitted beam while maintaining the intensity. The polarised beam was then focused by a lens (L3) with a focal length of 200 mm onto the back focal plane (BFP) of an oil-immersion TIRF objective (CFI Apo TIRF 100X, 1.49 NA, Nikon) through a non-polarising beam splitter (BS). The accurate positioning of the focused spot onto the BFP in the radial direction (x) was achieved with a z-axis motorised translation stage, leading to a collimated beam emerging from the objective at an incident angle ($\theta_{\rm{i}}$) proportional to the radial displacement of the focused spot from the optical axis of the objective. The incident collimated beam in the glass coverslip is totally reflected while producing an evanescent field at the glass-water interface (See Supplementary Figure S1). A custom-built coverslip chamber (See inset of Figure\,\ref{fig:1}) was designed to hold two 22-mm coverslips with a 5-mm gap between them. Such water spacing is wide enough for injecting medium and placing the glass microsphere near the surface of the coverslip. The reflected spot at the BFP was projected onto a plane conjugate to BFP of the TIRF objective through a Bertrand lens (L4) and a tube lens (L5) with focal lengths of 150 mm and 200 mm, respectively. A pinhole was placed at the conjugate plane to remove unwanted stray light and increase the signal-to-noise ratio. The spatially filtered beam that contains optoplasmonic signals was subsequently collimated and focused onto an amplified photodiode (PD) by relay lenses (L6 and L7). For fluorescence detection, both switchable mirror (SM) and L4 were replaced by an emission filter to allow quick alteration between optoplasmonic sensing and TIRF imaging. Fluorescence images were recorded by a back-illuminated EMCCD camera (iXon Ultra 888, Andor). The laser intensity hitting the coverslip was $\approx$32 $\rm{W/cm}^2$.

\subsection*{Experimental protocols}
In this work, we used two sets of DNA oligos (See Table 1 for sequences): (i) a 10-mer docking strand termed P1 and corresponding imager termed ImP1*D, and (ii) a 22-mer docking strand termed T22 and corresponding imager strand ImT22*D. The experimental procedure for both DNA-PAINT and optoplasmonic sensing are similar and consist of 3 main steps. First, gold nanorods were deposited onto a glass surface (either a coverslip in the case of DNA-PAINT, or microsphere resonator surface in the case of the optoplasmonic sensor) in an acidic aqueous suspension (pH $\approx$ 1.6, 1 pM GNRs) for 15 min. The sample chamber was then washed thrice to remove unbound GNRs. For DNA-PAINT, a clean coverslip was prefunctionalised with PLL-g-PEG (Su-Sos) to prevent nonspecific binding between the fluorophore and the coverslip. During the deposition of GNRs, clear binding step signals were observed via the optoplasmonic sensing approach (See Supplementary Information Figure S5b). As for the DNA-PAINT approach, the location  of GNRs could be visualised by the EMCCD camera owing to their photoluminescence. Second, the docking strands were immobilized onto the GNRs through a mercaptohexyl linker at their 5' end. The P1 docking strands were immobilised in a citrate buffer at pH $\approx$ 3 with 1 mM NaCl. The T22 docking strands were immobilised in 0.02\% wt/wt sodiumdodecylsulfate (SDS) solution at pH $\approx$ 3. Before adding the docking strands into the sample chamber, they were pre-mixed with a solution containing a reducing agent (\SI{10}{\micro L} of $10 \,\rm{mM}$  TCEP) to cleave the disulfide bonds and therefore enable efficient binding of the thiols to the GNRs. The docking strands were then injected into the sample chamber to reach a final concentration of \SI{1}{\micro M} and left to incubate for 30 min. Since the docking strands do not contain any fluorescent labels, the binding of the docking strands to the GNRs were not monitored via DNA-PAINT. In contrast, in the optoplasmonic sensor, step-like signals were observed upon binding of the docking strands to GNRs (See Supplementary Information Figure S6a). It is important to note here that only a subset of all docking strands attached to the GNRs provided step signals due to the variability in plasmonic enhancements at each binding site.\cite{eerqing2021comparing} Finally, the transient interactions between the docking and imager strands were monitored using both techniques. The chamber was washed three times to remove excess docking strands, and corresponding labelled imager strands were added to the sample chamber in an aqueous solution of 20 mM HEPES buffer at pH $\approx$ 7, with two different NaCl concentrations of 10 mM and 500 mM, respectively. All experiments were conducted at room temperature of around 295.7 K.

\subsection*{Single-Molecule Localisation Imaging via DNA-PAINT}

Fluorescence imaging was carried out on the dual sensing and imaging platform, with an effective pixel size of 130 nm. For DNA-PAINT measurements, 2000 frames per image were recorded at a frame rate of 25 Hz. The EMCCD readout rate was set to 30 MHz at 16-bit resolution and a pre-amplifier gain of 2. The electron multiplier gain was set to 750. Utilising the single-molecule localisation microscopy add-in (ThunderSTORM\cite{ovesny2014thunderstorm}) in Fiji (ImageJ), one can obtain the intensity time traces for single-molecule localisation on the GNRs. The image filtering was performed via Wavelet filter (B-Spline), where B-Spline order was set to 3 and scale to 2. The approximate localisation of molecules has been carried out via local maximum method, with 8-connected neighbourhoods and a peak intensity threshold set to the standard deviation of the 1st wavelet level. The sub-pixel localisation was achieved from point spread function estimation via integrated Gaussian fitting. The fitting radius was set to 3 pixels and the initial sigma set to 1.6 pixels. 

 \subsection*{Optoplasmonic Sensing of DNA Hybridisation}

The mechanism of WGM based optoplasmonic sensing is as follows. Optical WGMs are excited in a glass microsphere resonator by coupling the evanescent wave emerging from the surface of the glass coverslip. The external cavity laser is then scanned over a small bandwidth around the WGM resonance to record the resonance spectra as shown in the inset of Figure\,\ref{fig:1}d. Plasmonic nanorods are then attached to the surface of the WGM resonator using the protocols described above and their plasmon resonance is excited by the WGM. When molecules approach the near field of the GNRs, shifts in the resonance of the WGM are produced due to a local change in refractive index (the molecule is essentially polarised by the electric field which in turn causes a shift in wavelength $\Delta\lambda$) as shown by the red trace in the inset of Figure\,\ref{fig:1}d. By tracking the shift in resonance wavelength ($\Delta\lambda$) and the linewidth ($\Delta$ FWHM, $\Delta\kappa$), we can monitor real-time single-molecule interactions at/near the hotspots of GNRs. To extract $\Delta\lambda$ and $\Delta\kappa$ from the resonance spectra, a centroid fitting algorithm is applied. Custom MATLAB code was employed to detect peaks, remove background trends (i.e. detrending), and estimate event rates. A threshold of 3$\sigma$ is applied for signal detection. \cite{baaske2014single}

\section*{Results}

\begin{figure}[t]
\centering
\includegraphics[width=0.95\linewidth]{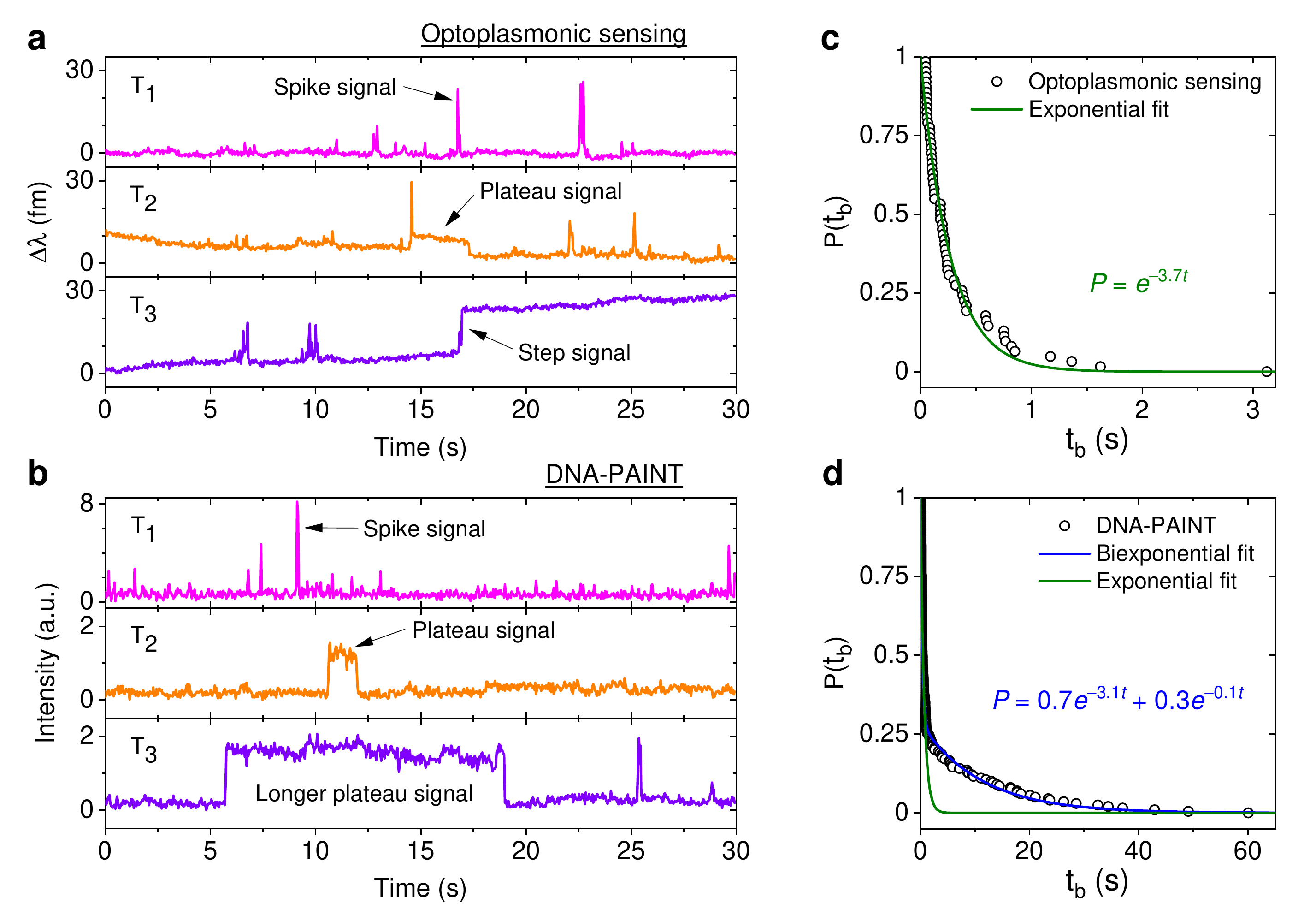}
\caption{Three types of DNA hybridisation signals. Signals obtained from (a) optoplasmonic sensor and (b) single-molecule imaging platform at different time intervals (T1,T2,T3). (c) Measured $t_{\rm{b}}$ from the P1 optoplasmonic data and its single exponential fit. The probabilities $P(t_{\rm{b}})$ represent a dissociation event has not taken place within an interval $t_{\rm{b}}$ .(d) $P(t_{\rm{b}})$ for the same P1 DNA sets measured by DNA-PAINT and their single (green) and bi-exponential (blue) fits.}
\label{fig:2}
\end{figure}

\subsection*{Signal Types and Dissociation Kinetics from Optoplasmonic and DNA-PAINT Measurements}
DNA hybridisation is characterised using both optoplasmonic sensing and DNA-PAINT. Two sets of DNA docking strands (P1/T22) and the corresponding imager strands (ImP1*D/ImT22*D) were utilised in both experiments, their sequences are detailed in Table \ref{tab:T1}. Figure \,\ref{fig:2} displays the typical signal type from a data trace. It can be seen that most signals along these time traces are spike-like with a short dwell time regardless of the chosen measurement technique. Prolonged plateau-like signals were, however, also observed as in the central traces (T2 time interval) of Figure\,\ref{fig:2}a, b. Positive step-like signals with no corresponding falling-edge were observed in the case of the optoplasmonic sensor (T3, Figure\,\ref{fig:2}a). These signals most likely correspond to the permanent occupation of a docking strand by an imager strand. On the contrary, no such step signals were observed in the DNA-PAINT experiments. In this case, longer plateau-like signals (T3, Figure\,\ref{fig:2}b) were observed instead.  The spike-like signals and transient plateau signals in both techniques (i.e. Signals in Figure\,\ref{fig:2}a, b's T1,T2) correspond to reversible / transient interactions between the imager and docking strands. The step signal is only observed for the optoplasmonic sensing method, correlating to the permanent (explicitly, within the timescale of the experiment) hybridisation of the complementary strand\cite{baaske2014single}. On the other hand, DNA-PAINT only provides plateau-like signals with a longer duration (T3, Figure\,\ref{fig:2}b). We hypothesise that these long-duration, plateau-like signals also correspond to permanent hybridisation events. The falling edge is thus likely due to photobleaching of the dye after a permanent hybridisation event rather than from dissociation of the DNA strands.

We further analysed the signals to extract the dwell time of events $t_{\rm{b}}$ in case of spike-like and plateau-like signals, the arrival time between consecutive events $t_{\rm{a}}$.
Figure\,\ref{fig:2}c, d plots the survivor function for the signal dwell time $t_{\rm{b}}$ measured with both techniques. It can be seen that the data from the optoplasmonic sensor (Figure\,\ref{fig:2}c) closely matches a single exponential fit, corresponding to a single Poisson process. The dwell rate estimated from a single exponential fit is $k_{\rm{off}}^{\rm{OP}}= 3.7\pm 0.1~\rm{s}^{-1}$. However, in the case of the DNA-PAINT measurement (Figure\,\ref{fig:2}c), the distribution of dwell times largely deviates from a single exponential. A bi-exponential model provides a better fit, where two rates $k_{\rm{off,1}}^{\rm{PAINT}}= 3.1\pm 0.1~\rm{s}^{-1}$ and $k_{\rm{off,2}}^{\rm{PAINT}} = 0.10\pm 0.06~\rm{s}^{-1}$ are obtained. The rates $k_{\rm{off}}^{\rm{OP}}$ and $k_{\rm{off,1}}^{\rm{PAINT}}$ have similar values which correspond to the binding dwell time of the DNA strands. The rate $k_{\rm{off,2}}^{\rm{PAINT}}$, however, likely corresponds to the average photobleaching time of the dye molecule under experimental conditions. 

\begin{figure}[t]
\centering
\includegraphics[width=0.85\linewidth]{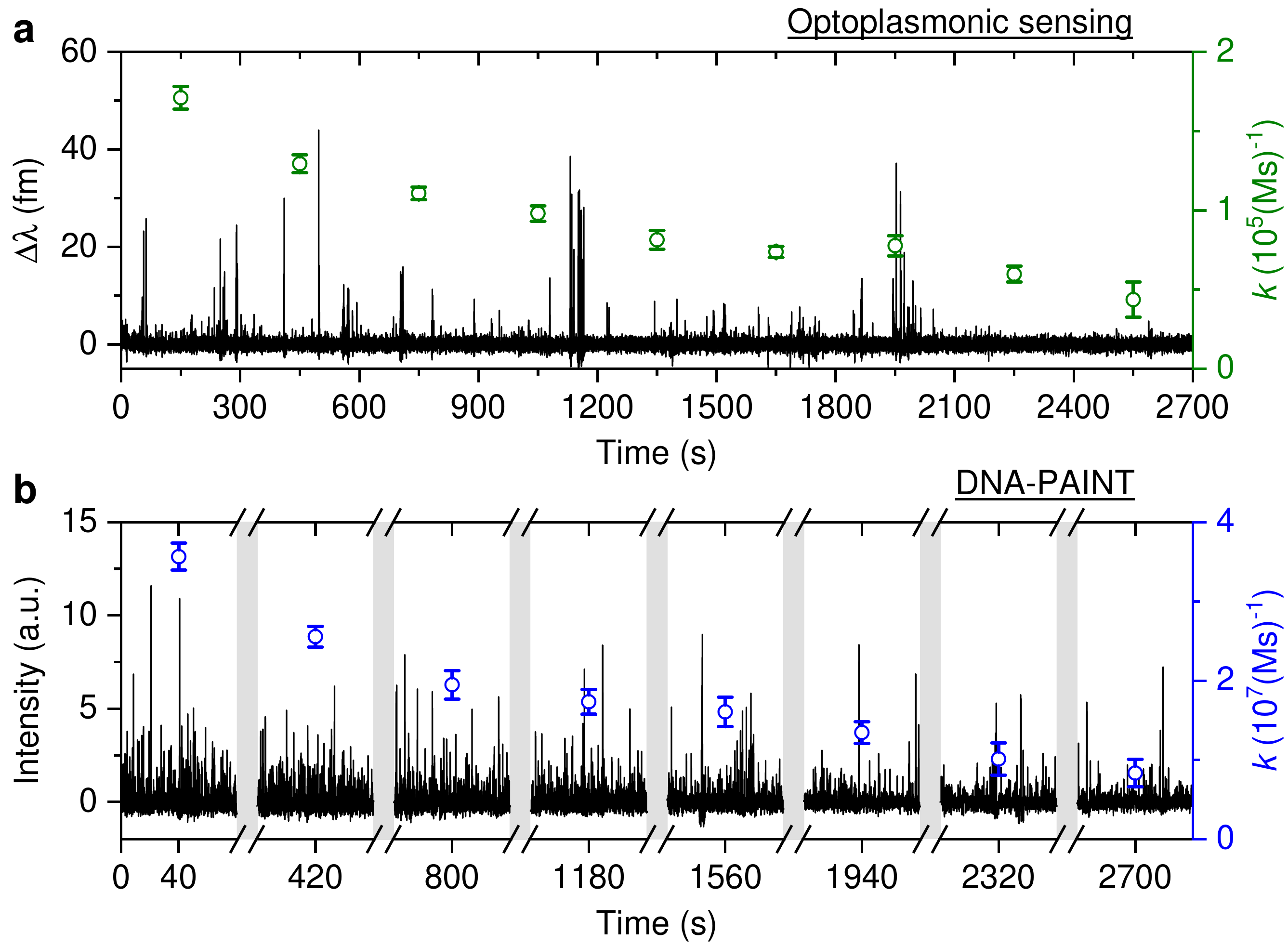}
\caption{T22 DNA association rates change measured over time. Experiments are carried out with optoplasmonic sensing and DNA-PAINT approaches with 10mM NaCl buffer. (a) Detrended optoplasmonic sensing trace for \SI{1}{\micro M} ImT22*D and association rate (green circles) calculated every 300 s. (b) DNA-PAINT fluorescence time trace with 20 nM ImT22*D, with association rates (blue circles) measured every 5 minutes within time intervals of 80 s. The uncertainties of the association rates are extracted from exponential fits}
\label{fig:3}
\end{figure}

\subsection*{Measuring Association Rate Change Over Time}

The signal patterns and dwell time analysis above indicate that the spike-like signals arise from transient DNA hybridisation. The step-like signals for the optoplasmonic sensor and the long plateau-like signals in DNA-PAINT likely arise from hybridisation / blocking of docking DNA. Furthermore, the analysis of dwell times indicates that longer plateau-like signals in DNA-PAINT correspond to a permanent blocking of a DNA docking strand by an imager strand and subsequent photobleaching of the dye attached to the imager. 

To study the anomalous behaviour of DNA hybridisation kinetics on GNRs as was previously reported\cite{taylor2018all, zhang2018single}, we performed long timescale measurements ($\sim$ 45 mins) of DNA hybridisation. To ensure the observation was not sequence dependent, we utilised the second set of DNA strands: T22 and the corresponding imager (See Table\,\ref{tab:T1}). The experimental protocol for DNA functionalisation and imaging is the same as in the previous trial.
Figure\,\ref{fig:3} (black traces) shows the long time traces obtained from both techniques. The concentration of the imager strands for optoplasmonic sensing and DNA-PAINT measurements are \SI{1}{\micro M} and 20 nM, respectively. Different concentrations are used for optoplasmonic and DNA-PAINT measurements due to the different distributions of detectable docking sites\cite{eerqing2021comparing}. In the case of DNA-PAINT (Figure\,\ref{fig:3}b), 5 min of idle time in the measurement is delineated by break lines.

The plots in Figure\,\ref{fig:3} show the anomalous behaviour of a decrease in the number of events over time for both techniques. To identify the decreasing rate of events $k$ (association rates corresponding to arrival times $t_{\rm{a}}$), we estimated the event rate $k$ over multiple consecutive short intervals. The interval duration was 300 s for measurements using the optoplasmonic sensor and 80 s for measurements using DNA-PAINT. These measurement intervals are a practical compromise between obtaining good time resolution and good exponential fits for determining the average association rates in the 300 s / 80 s measurement intervals.  The association rates are obtained in these intervals by fitting single decaying exponentials to the time between consecutive arrivals $t_{\rm{a}}$. The rates $k$ are plotted on the time traces in Figure\,\ref{fig:3}a (green circles with fit errors) and Figure\,\ref{fig:3}b (blue circles with fit errors). We can clearly see that the association rates decrease consistently over time. The observed reduction in association rate for both platforms clearly suggests that, over time, the number of docking strands available for observing transient hybridisation events is decreasing. According to previous literature\cite{zhang2018single,lin20203d}, this phenomenon is not observed when DNA-PAINT experiments are performed on biomolecules bound to glass / polystyrene substrates\cite{lutz2018versatile,lin20203d}.

\begin{figure}[t]
\centering
\includegraphics[width=0.85\linewidth]{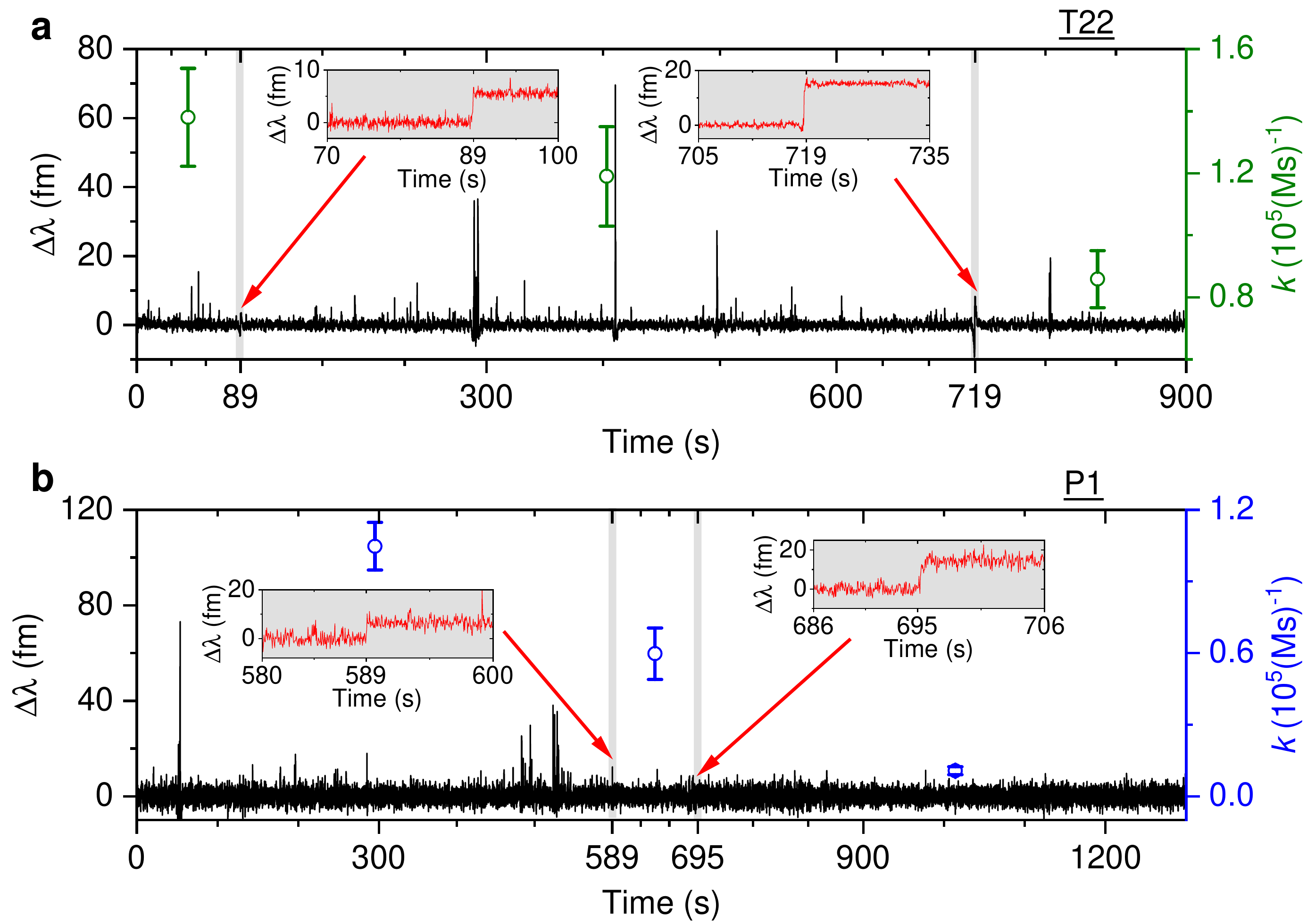}
\caption{Example of detrended wavelength shift traces (black). An optoplasmonic sensing technique is applied to measure \SI{1}{\micro M} of (a) ImT22*D and (b) ImP1*D imager strands. The grey shaded area displays the time period when a step signal is detected, a zoomed-in view of the step signals is shown in the inset images (red traces). The association rate is calculated for each section before and after the step signals (green circles for T22 and blue circles for P1 DNA sets).}
\label{fig:4}
\end{figure}

We then study the relationship between change in the association rate with respect to the occurrence of detected step signals. Optoplasmonic sensing experiments were performed for both T22 and P1 DNA sets. In particular, \SI{1}{\micro M} of ImT22*D and ImP1*D imager strands are used to observe the transient interaction. Figure\, \ref{fig:4} shows the typical time trace for wavelength change in the context of T22 and P1 DNA sets. In both time periods we observe two step signals, wherein we analysed the association rates before and after the step signals. The time intervals of detected step signals are marked in the grey region in  Figure\,\ref{fig:4} and a zoomed-in view of the step signals (red trace) is shown in the inset images. The association rates for each interval is then plotted (green circles for T22 and blue for P1 sets). Obvious association rate drops after each step are shown in both T22 and P1 experiments. Especially for P1 experiments depicted in Figure\,\ref{fig:4}b, one can clearly see that after two step signals, the association rate has decreased from $(1.0\pm 0.1) \times 10^5(\rm{M\cdot s})^{-1} $ to $(1.1\pm 0.2)\times 10^4(\rm{M\cdot s})^{-1}$. This observation indicates that the drop in anomalous DNA hybridisation rate is correlated to the step signals. As discussed above, a positive step signal refers to a binding event. There is no corresponding falling edge likely due to the imager strands irreversibly hybridised to the docking strands. This would then prohibit the docking strand from further interactions and hence result in a decreased association rate.

\subsection*{Imager Concentration Dependence for Prolonged Measurements}

\begin{figure}[t]
\centering
\includegraphics[width=1\linewidth]{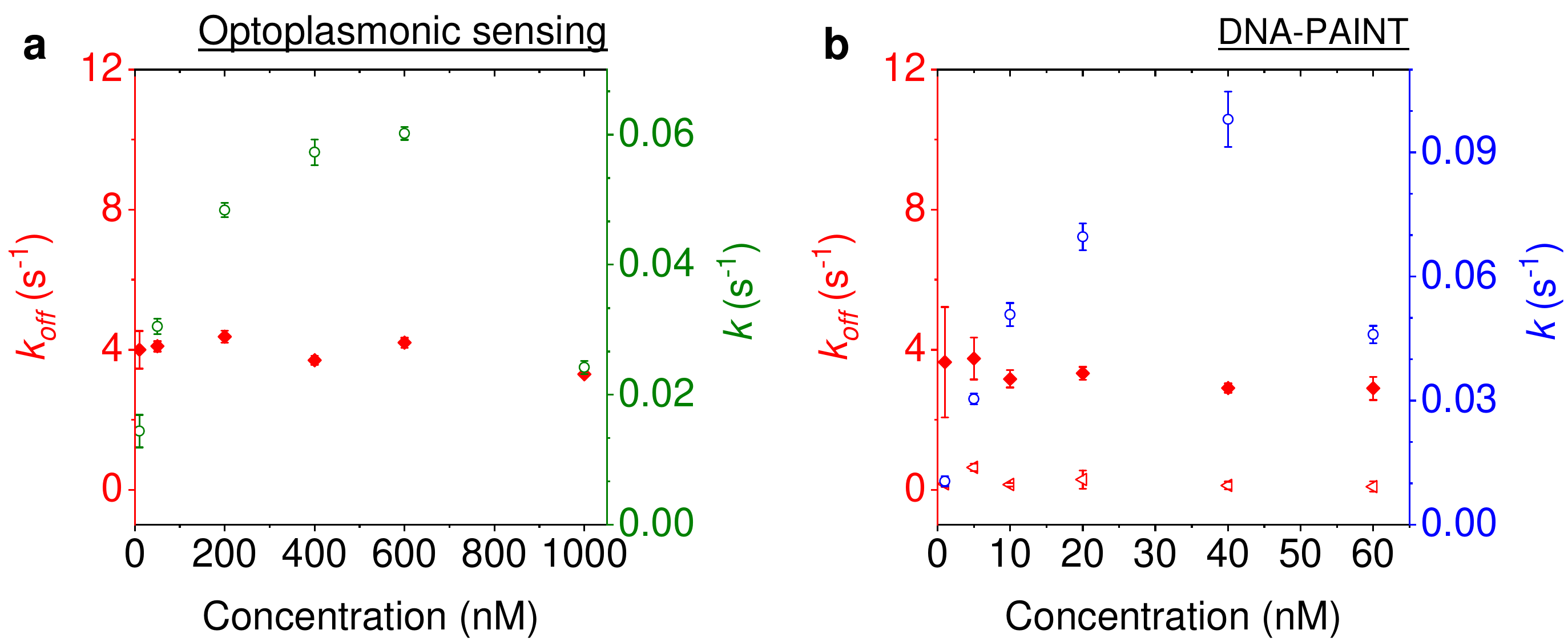}
\caption{Kinetics characterised for a P1 DNA set.  The experimental data is acquired with (a) the optoplasmonic sensor and (b) DNA-PAINT.  Association rates are plotted for optoplasmonic data (green circles)  and DNA-PAINT data (blue circles). Dissociation rates are extracted from fitting of single exponentials for  optoplasmonic sensing (red dots in a) and biexponentials for DNA-PAINT (red dots and triangles in b). Both approaches are measured for 6 different imager strand concentrations and each concentration is measured over 30 min.}
\label{fig:5}
\end{figure}

We next studied the concentration dependence of the arrival and dwell times by increasing the concentration of the imager strand, in 6 steps, for both techniques. The measurements at each concentration step were performed over 30 min. The dissociation rate $k_{\rm{off}}$ and the association rate $k$ for the DNA strand P1 and its complementary imager are plotted in Figure\, \ref{fig:5}a and Figure\, \ref{fig:5}b for the optoplasmonic sensor and DNA-PAINT, respectively. The dissociation rates $k_{\rm{off}}^{\rm{OP}}$ for the optopalsmonic sensor and $k_{\rm{off}}^{\rm{PAINT}}$ for DNA-PAINT are estimated from a single- and bi-exponential fit to the survivor distribution of dwell times $t_{\rm{b}}^{\rm{OP}}$ and $t_{\rm{b}}^{\rm{PAINT}}$, respectively. The association rates $k$ for both techniques are obtained from a single exponential fit to the survivor distribution of time between events $t_{\rm{a}}$.  

The plots show that the off-rates $k_{\rm{off}}$ provided by both techniques are relatively constant for all imager concentrations. The average off-rate estimated from the optoplasmonic measurements is $k^{\rm{OP}}_{\rm{off}} = 3.9\pm 0.3\,\rm{~s^{-1}}$. For the DNA-PAINT measurements, the average off-rates estimates are $k_{\rm{off,1}}^{\rm{PAINT}}=3.3\pm 0.7~\rm{s}^{-1}$ and $k_{\rm{off,2}}^{\rm{PAINT}}=0.2 \pm 0.1~\rm{s}^{-1}$. Here, the values of $k^{\rm{OP}}_{\rm{off}}$ and $k_{\rm{off,1}}^{\rm{PAINT}}$ are within error and in line with previous work \cite{eerqing2021comparing}. These rates thus correspond to the actual hybridisation dwell times of the P1 and imager strands. The rate $k_{\rm{off,2}}^{\rm{PAINT}}$ likely corresponds to the average photobleaching time of the dye (DY782) of $5 \pm 0.5\, \rm{s}$ under the experimental conditions.

On the other hand, the association rates $k$ increase with higher imager concentrations. The increase in $k$ for both techniques is not linear with imager concentration as would be expected for single-molecule interactions. The estimated $k$ gradually saturates with increasing imager concentrations and even drops significantly, as is seen at 1000 nM for the optoplasmonic sensor and 60 nM for DNA-PAINT. This deviation from linearity may arise from an increased probability of docking strands being permanently occupied for higher imager concentrations. The drop in event rate k occurs at different concentrations, which is attributed to the discrepancy between effective sensitivities of DNA-PAINT and optoplasmonic sensing. In the case of optoplasmonic sensing, the docking strands that provide signals are the ones located on the hotspots. The hotspots are distributed towards the tip areas of the gold nanorods, accounting for around 22\% of the total surface area of the gold nanorod. Only up to 22\% of the docking strands therefore contribute signals in the optoplasmonic sensing setup. On the contrary, all docking strands on the surface of the gold nanorod contribute to DNA-PAINT signals. Furthermore, these two different approaches show different signal-to-noise ratios, both of which provide a plausible picture for the discrepancies we observed. A detailed discussion of the discrepancy in measuring association rate of DNA hybridisation on both techniques is provided in our previous work \cite{eerqing2021comparing}.

\subsection*{Control Measurements and Investigation of Potential Causes}

The signals observed in both DNA-PAINT and optoplasmonic sensing imply an increasing number of unavailable docking strands over time. Based on the step signals, a natural hypothesis would be that permanent hybridisation block the docking strands within the time scale of the experiment. Alternatively, the docking strands could be cleaved due to hot electrons generated from the plasmonic nanorods. Control experiments are performed to discern the causes of the reduction in association rates over time.

In 2018, Sabrina \textit{et al}.\cite{simoncelli2018nanoscale} demonstrated that a high-power pulsed laser could cleave  $Au-S$ bonds on the surface of the GNRs, wherein the cleaved surface can be refunctionalised with different docking strands. There is, however, no evidence for the low intensities (32 $\rm{W/cm}^2$) in our experiment causing the cleaving of $Au-S$ bonds. Nonetheless, to prove that the $Au-S$ bonds were intact, we performed control measurement using samples where the association rate already dropped to near zero. We washed the samples (i.e. coverslips immobilised with gold nanorods, functionalised with docking DNA and left to interact with imager strands for $>$ 6 hours) thrice with Milli-Q water to remove excess DNA. We then performed the protocols for immobilizing new thiolated docking DNA (See Materials and Methods) for 1 hour to fully saturate any available binding sites on the GNR (assuming laser cleaving of previously attached docking DNA). In the optoplasmonic sensing method, we do not observe any docking DNA binding steps in this process, indicating no available sites on the GNR surface (See Supplementary Figure S2a). Finally, we added \SI{2}{\micro M} of imager DNA in the case of optoplasmonic sensing and 40 nM of imager strands in the case of DNA-PAINT measurements. As seen from the time traces obtained (See Supplementary Figure S2b, c), we observe no restoration of association rate from both platforms, indicating that new docking strands could not be immobilised. We hence conclude that there is no laser-induced cleaving of the initially functionalised docking DNA in our experiments. The formation and dissociation of $Au-S$ bonds is known to be an equilibrium process; however, we did not observe single-molecule signals associated with ligand exchange between anomalously hybridized docking strands with fresh imagers. This indicates that the time scale for any ligand exchange exceeds our measurement duration of 45 min. Which matches with the literature \cite{woehrle2005thiol} that, for long thiols like thiolated DNA strands, the exchange time is normally over 12 hours.

\begin{figure}[t]
\centering

\includegraphics[width=0.95\linewidth]{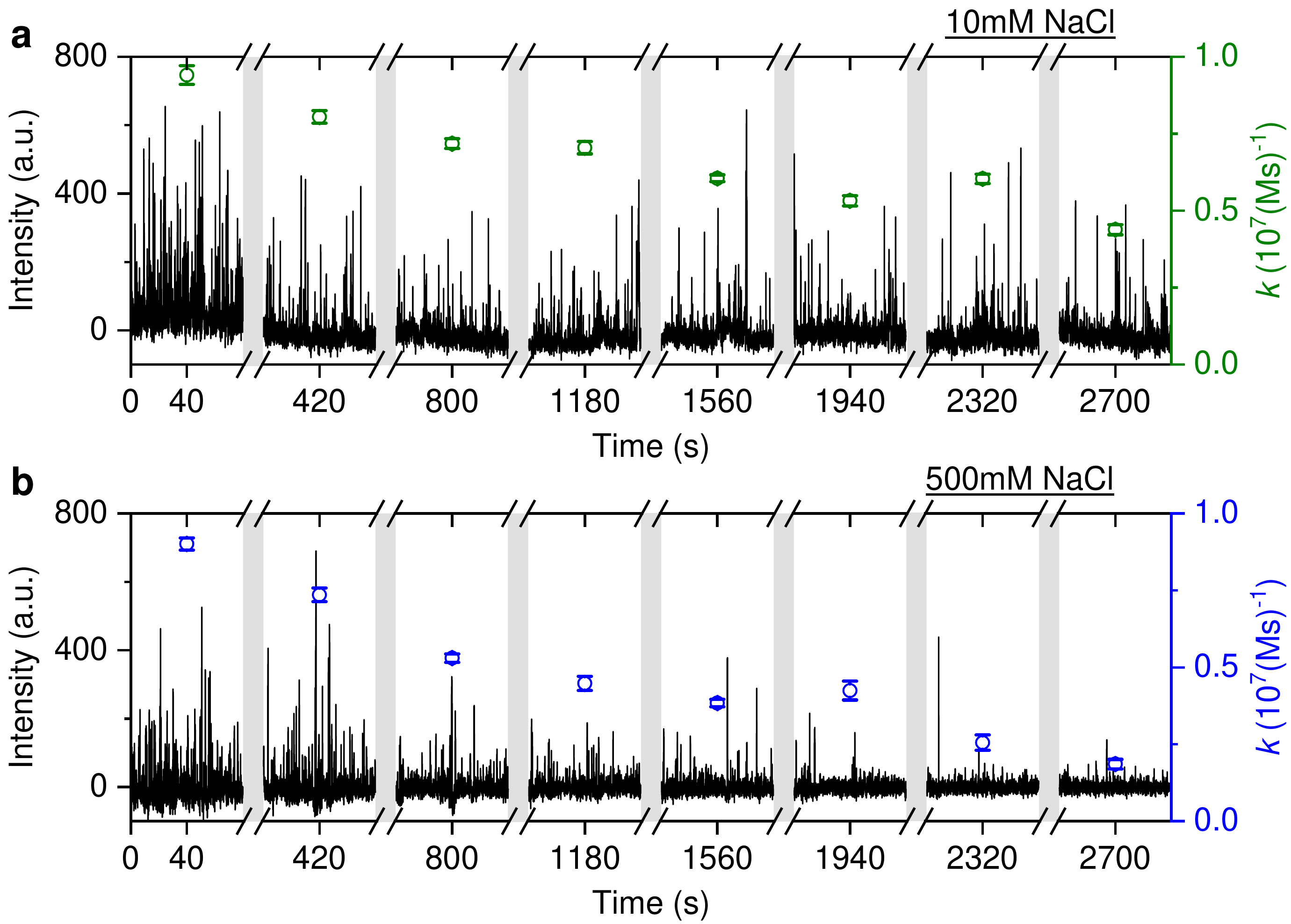} 
\caption{DNA-PAINT measurements with 40 nM ImP1*D. The experiment is conducted with (a) 10 mM NaCl HEPES buffer and (b) 500 mM NaCl HEPES buffer. The association rates are plotted accordingly, with green circles for 10 mM NaCl HEPES buffer and blue circles for 500 mM NaCl HEPES buffer.}
\label{fig:6}
\end{figure}

This additional control suggests that the observed anomalous hybridisation behaviour cannot be explained by hot electron cleaving. Given the steps signals demonstrated in the optoplasmonic sensing experiments, we can conclude that the anomalous DNA hybridisation originated from the permanent hybridisation of imager strands. To further investigate the mechanism of this anomalous hybridisation, two sets of experiments are then carried out. The first set of experiments aims to explore the influence of different electrolyte concentrations, and the second utilises external heating to inspect whether the anomalously hybridised DNA melts at a higher temperature. 

It is well known that adding electrolytes in a solution can affect the DNA melting temperature\cite{gruenwedel1969salt}. As we concluded the observed anomalous behaviour resulted from permanent hybridisation, it is expected that this will also depend on electrolyte concentration. To this end, we monitored the association rate change for different ionic strengths in the buffer solution. In particular, a solution with 40 nM ImP1*D imagers is used to conduct DNA-PAINT measurements under 10 mM and 500 mM NaCl concentrations in the buffer. Both experiments are carried out for over 45 min and the intensity time trace is plotted in Figure\,\ref{fig:6} (black trace). The association rate is calculated and plotted as green circles for 10 mM NaCl HEPES buffer and blue circles for 500 mM NaCl HEPES buffer. To properly compare the rates decreasing under different NaCl concentrations, we selected GNRs that have similar starting association rates corresponding to the 10 mM and 500 mM NaCl concentrations. As shown in Figure\,\ref{fig:6}, the association rate for the first 80 s is calculated to be $(9.4\pm0.3) \times 10^6 (\rm{M\cdot s})^{-1}$ for the 10 mM NaCl HEPES buffer and $(9.0\pm0.2) \times 10^6 (\rm{M\cdot s})^{-1}$ for the 500 mM NaCl HEPES buffer, respectively. In both time traces we observe a similar decrease in association rates. In the case of the 500 mM NaCl HEPES buffer, however, the association rate decreases much faster. After 45 min the remaining association rate for the 10 mM NaCl HEPES buffer is $(4.4\pm0.2)\times 10^6 (\rm{M\cdot s})^{-1}$, while for the 500 mM NaCl condition a much lower rate of $(1.8\pm0.2) \times 10^6 (\rm{M\cdot s})^{-1}$ is found. The observed phenomenon further matches the permanent hybridisation hypothesis. Higher ionic strength can be induced in the solution with increased NaCl concentration, which lowers the charge barrier and provides better accessibility for the docking strands. The permanent hybridisation process is thus accelerated due to increased hybridisation efficiency.

All experiments described above are conducted at $\sim$ 295 K room temperature. Given that plasmonic nanorods confine light to a nanoscale region, one would expect a higher temperature on the GNR surface than the surrounding host medium. The higher temperature on the GNR surface could potentially cause an equilibrium perturbation in DNA hybridisation. To estimate the temperature change caused by plasmonic heating, we calculated the temperature increase with a laser intensity of 80 $\rm{W/cm}^2$ - the maximum intensity delivered to the glass coverslip. According to simulations (Detailed in Supplementary Figure S3), the maximum GNR surface temperature is only increased by 0.1 K at the GNR absorption peak wavelength of 755 nm. Considering that a GNR is placed onto the surface of the glass coverslip under 785-nm TIRF excitation which exhibits a near-field enhancement factor of 5 (Supplementary Figure S1), the estimated temperature gradient near the GNR surface is not higher than 0.5 K with a laser intensity of 32 $\rm{W/cm}^2$. This temperature increase is too small to cause a significant change in the hybridisation equilibrium. In experiment, no restored rate is observed when increasing the laser illumination power from 32 $\rm{W/cm}^2$ to 80 $\rm{W/cm}^2$. 

To examine the effect of temperature increase on the melting of anomalous DNA hybridisation, we increased the temperature of the buffer solution externally  (See Supplementary Figure S4). In this scenario, DNA-PAINT experiments are conducted under different temperature conditions. We again choose P1 DNA samples where the association rate already decreased to near zero, wherein the sample is acquired using the same protocol described above. The procedure is as below. First, the chamber is filled with HEPES buffer heated to a temperature of $\sim$ 333 K for 30 min, to enable potential melting of double-stranded DNA. Then, the heated solution is removed and sample chamber is rinsed thrice to remove possibly unzipped imager strands. Finally, a HEPES buffer with 10mM NaCl and containing 40 nM of imager is added to the sample chamber to perform DNA-PAINT experiments.

Supplementary Figure S4a shows the DNA-PAINT signals obtained when measurements are performed in a buffer with 40 nM imager concentration heated to $\sim$ 333 K. At this higher temperature, the association rates of single-strand DNA are higher\cite{zhang2018predicting,straus1972temperature}. Nonetheless, we do not observe many events showing that the association rates are not restored at this temperature. To exclude the possibility that the higher dissociation rate at 333 K is faster than our time-resolution for DNA-PAINT, we performed measurements in a room temperature buffer containing 40nM imager. As shown by the time traces in Supplementary Figure S4b, we observe no restoration of association rates. Similar experiments at a higher temperature of $\sim$ 353 K also show no restoration of association rates (Supplementary Figure S4c, d). These findings suggest that the anomalous hybridisation of DNA attached to gold nanoparticles are irreversible and highly stable (up to $\sim$ 353 K).

\section*{Discussion}

Here, we have established a hybrid platform (SIMOPS) capable of performing both fluorescence based single-molecule localization imaging (DNA-PAINT) and refractive index based optoplasmonic sensing. We have utilized this platform to study the anomalous hybridisation kinetics of short single-stranded DNA (ssDNA) oligomers attached to the surface of gold nanoparticles. The SIMOPS platform provides us with time traces of different signal types corresponding to the hybridisation kinetics of ssDNA. Both DNA-PAINT and optoplasmonic sensing methods provide spike-like (duration of a few hundred milliseconds) and short plateau signals (duration of a few seconds) that are corresponding to the transient hybridisation of the ssDNA. The optoplasmonic sensor also provides positive step-like signals with no corresponding falling edge indicating permanent hybridisation of the ssDNA. A single kinetic off-rate of $k_{\rm{off}}^{\rm{OP}}= 3.7\pm 0.1~\rm{s}^{-1}$ is reported from the optoplasmonic sensing approach. DNA-PAINT on the other hand provides long plateau-like signals with both positive and negative steps separated by tens of seconds. We have shown that these signals likely correspond to the permanent hybridisation of the ssDNA and subsequent bleaching of the imagers. DNA-PAINT hence provides two kinetic off-rates with values ($k_{\rm{off,1}}^{\rm{PAINT}}=3.3\pm 0.7$ s) and ($k_{\rm{off,2}}^{\rm{PAINT}}=0.2\pm0.1$ s). The magnitudes of $k_{\rm{off}}^{\rm{OP}}$ and $k_{\rm{off,1}}^{\rm{PAINT}}$ are consistent with predictions for DNA melting temperature specific to the oligonucleotides and buffer conditions\cite{zhang2018predicting} and with measurements performed using similar DNA sequences at room temperature without local plasmonic heating\cite{jungmann2010single}. The magnitude of $k_{\rm{off,2}}^{\rm{PAINT}}$ associated with photobleaching of the DY782 at the laser intensity of  $\sim$ 32$\rm{W/cm}^2$ is comparable to that reported in the literature \cite{wustner2014photobleaching,tsunoyama2018super}. 

Measurements of DNA hybridisation kinetics over a long time ($\sim$ 45 mins) show that the association rate gradually decreases with time as measured by both DNA-PAINT and the optoplasmonic techniques. We show that the decrease in association rate is correlated to the step-like signals observed via the optoplasmonic sensing, indicating a permanent occupation of 'docking' DNA strands attached to the GNR surface. We found that the association rate increased non-linearly with increasing imager concentration over prolonged time, which even decreased at very high imager concentrations. Our data demonstrates the decrease in association rate behaviour can be accelerated by increasing the ionic strength. These observation further bolstering the hypothesis of permanent hybridisation of the 'docking' and 'imager' strands on the surface of the GNRs.

In control measurements we attach more 'docking' DNA to GNRs where the association rates dropped to near-zero, show that no new 'docking' could be attached to the GNRs. This observation is inconsistent with previous reports of hot electron-mediated cleavage of $Au-S$ bonds of the docking strands \cite{simoncelli2018nanoscale}. Additionally, we observe no falling (negative) step-like signals in the optoplasmonic sensor data also indicating that there is no hot-electron cleaving of the $Au-S$ bonds. Instead, our data and control measurements suggest a permanent hybridisation of the 'docking' and 'imager' strands. 

Further, control measurements with increasing the temperature of the buffer show that the anomalous DNA hybridisation is stable up to at least $\sim 353$ K. One possible explanation for anomalous permanent hybridisation of DNA strands to complementary strands on plasmonic gold nanoparticles is DNA inter-strand cross-linking. \cite{rozelle2021dna} The DNA inter-strand cross-links could be caused by the oxidative chemical environment found at/near the plasmonic hotspots of the gold nanoparticles, i.e. the sites of DNA hybridisation at which single-molecule signals are generated. More specifically, singlet oxygen and radical oxygen species could be locally generated by the light-metal interaction \cite{carrasco2020plasmonic}. Various other photochemical processes could result in DNA mutations leading to anomalous permanent hybridisation events. Further research into different DNA sequences are needed to fully understand this mechanism.

No such anomalous hybridisation signals, with prolonged dwell times that lie well outside of the expected dwell time distributions for these oligonucleotides, have been reported before\cite{peterson2016single}. Our observations contradict the prediction of well-established theoretical models that consider the thermodynamics of base-pairing interactions of DNA nucleotides\cite{peterson2018identification,peterson2002hybridization,xu2017real}. The anomalous hybridisation events reported here have not been detected before because established single-molecule techniques, such as those that use fluorescently-labelled oligomers, have been unable to track permanent hybridisation events over prolonged times due to photobleaching. On the other hand, Surface Plasmon Resonance (SPR) based sensors do not strictly show single-molecule sensitivity. They are only capable of observing wavelength shifts by bulk loading, remaining insensitive to transient DNA hybridisation events. It is thus challenging to observe anomalous hybridisation events without relying on additional signal transduction pathways or a direct single-molecule technique. The proposed SIMOPS single-molecule platform is capable of discerning transient from permanent interactions of short oligonucleotides (i.e. localised on the plasmonic nanoparticles) over measurement times of up to several hours. Our dual DNA-PAINT and optoplasmonic sensing platform will enable further detailed investigations of DNA mutations in real-time and at the single-molecule level. This technique paves the way to real-time observation of single-molecule events common between two techniques to obtain more information content on single-molecule processes such as anomalous DNA hybridisation. SIMOPS could be further used for single-molecule experiments that reveal the occurrence of specific mutations. Such studies can differentiate the types of mutation and their respective kinetics with the use of proteins that selectively bind to specific mutations on a DNA strand. 

Our observations have implications for the use of localised plasmon resonance-based DNA sensors which affect the DNA interaction kinetics. Consequently, permanent DNA hybridisation results from prolonged light exposure at moderate light intensities. In single-molecule studies, as was shown here, such permanent hybridisation effects have to be carefully taken into account when interpreting single-molecule measurements. The study of the origin of anomalous DNA hybridisation signals can provide novel insights into the chemistry and photochemistry of DNA on plasmonic nanoparticle surfaces. Such studies can uncover yet-undetected pathways for introducing DNA mutations by interaction with light. Our studies also have implications for light-based in vivo therapies that use plasmonic nanoparticles, e.g. damage to tumour cells in the context of cancer treatments. In addition, recent studies have shown that DNA can work as a conductive material and transfer hot carriers generated along the plasmonic nanoparticles. Our platform is, therefore, ideal to study DNA hot electron charge transfer reactions that could trigger DNA interstrand crosslinking at a single-molecule level\cite{kogikoski2021spatial,rozelle2021dna,liu2020plasmonic,semlow2022hmces}.

\section*{Acknowledgments}
N.E. acknowledges funding from EPSRC Centre for Doctoral Training (CDT) in Metamaterials (XM2) (EP/L015331/1). N.E., H.-Y.W., S.S. and F.V. acknowledge funding from EPSRC EP/T002875/1. 
\section*{author contribution}

N.E conceived the idea, performed the experiments, and analysed data. H.-Y.W. conceived the original idea of TIRF-based excitation of the WGMs, designed and developed the optical setup, performed the numerical simulations, and contributed to figure plotting. N.E. and S.S. wrote the manuscript. S.S. guided the experiments and wrote the MATLAB scripts for analysis of optoplasmonic sensor data. S.V. trained N.E. in conducting optoplasmonic sensing experiments and helped with manuscript writing. F.V. supervised the project. All authors commented on the manuscript.

\section*{Supporting Information Available:}

The Supporting Information is available free of charge at:

Principle of TIRF-Based Excitation of the WGMs; Control Measurements by Refunctionalising the Docking Strands; Simulation of Plasmonic Heating; Control Measurements by External Heating; Background Measurements of DNA-PAINT and Optoplasmonic Sensing Systems; Visualising Real-Time DNA Interactions via Optoplasmonic Sensing; Example of Single-Molecule Localisation
\section*{Data availability}

The data that support the findings of this study are available from the corresponding author upon reasonable request.

\section*{conflict of Interest}

Authors declare no conflict of interests. 

\bibliography{references.bib}

\end{document}